\documentclass[journal=langd5,manuscript=article]{achemso}
\usepackage[T1]{fontenc}
\usepackage[nointegrals]{wasysym}
\usepackage{graphicx,amsmath,amssymb}
\usepackage{siunitx}
\usepackage{chemformula}
\usepackage{color,soul}
\newcommand{\Rey}{\mathrm{Re}}
\newcommand{\St}{\mathrm{St}}

\author{Yudong Li}
\author{Song-Chuan Zhao}
\affiliation{State Key Laboratory for Strength and Vibration of Mechanical Structures, School of Aerospace Engineering, Xi'an Jiaotong University, Xi'an 710049, China}
\email{songchuan.zhao@outlook.com}

\title{Freezing delay of a drop impacting on a monolayer of cold grains}

\begin{document}

\begin{tocentry}
%
%
%
%
%
\includegraphics{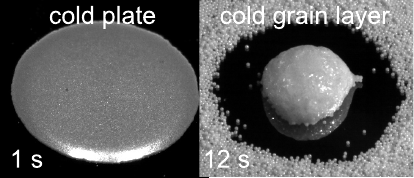}
\end{tocentry}
	
	\begin{abstract}
		We investigate a subfreezing droplet impact scenario in a low-humidity environment, where the target is a cold granular monolayer. When the undercooling degree of targets passes a threshold, such a granular layer effectively postpones the bulk freezing time of the droplet in comparison with the impact on the bare substrate underneath. In this case, the retraction of the droplet after impact reduces the contact area with the cold substrate, even though both the grains and the substrate are wettable to the liquid. We find that the significant changes in the dynamic behavior are triggered by freezing the liquid that wets the pores. Owing to the small dimension of the pores, the freezing process is rapid enough to match the dynamics over the droplet dimension. In certain circumstances, the rapid freezing may even stop liquid penetration and shed icing from the underneath surface.
	\end{abstract}

 	\section{Introduction}
 	Subfreezing droplet impact is relevant to various artificial and natural processes, such as aircraft icing, freezing rain hazards, and spray coating technology. In recent years, there has been a growing focus on anti-icing surface design~\cite{Kreder2016, Schutzius2015}. The superhydrophobic microstructure is commonly implemented to reduce the contact time of the droplet with the surface and to postpone its bulk freezing. Altogether, the impacting droplet is expected to be removed from the surface readily, for instance, by gravity. While there have been reports of freezing delay on the repellent coating~\cite{Jung2011,Guo2012,Alizadeh2012},  the potential for failures arises from liquid penetration through the microstructure~\cite{Maitra2014} and enhanced adhesion of the solidified residual~\cite{Varanasi2010,Kulinich2011,Chen2012}. Another challenge is the unavoidable random nucleation at high subfreezing degrees~\cite{Wang2019}, which impedes the superhydrophobicity and results in partial or complete adhesion~\cite{Schremb2017,Zhang2020}. Considering the complexity and variety of icing conditions, a unifying strategy of passive icephobicity is challenging.

	Recently, the shielding of icing was reported for droplet impact on a subfreezing sand pile~\cite{Zhao2024}. We suggest the same mechanism might apply to a monolayer of cold particles. 
	This paper explores the monolayer configuration in experiments, shedding light on the relative significance between the granular layer and the underlying substrate. Our findings confirm that a granular monolayer delays bulk freezing and, under specific conditions, can even detach the icing formed by an impacting drop onto the underlying substrate. In our experiments, a hexadecane drop of constant diameter ($D_0=2\pm 0.1\si{mm}$) at room temperature ($\approx 24\si{\celsius}$) impacts on a surface covered by a granular layer in a nitrogen chamber. The granular layer and the underneath silicone wafer are maintained at an undercooling degree $\Delta T$ below the melting point of the liquid before the impact. We use polydisperse ceramic (Yttrium stabilized Zirconium Oxide) spheres of mean diameters $ d=104,\, 172\si{\micro\meter} $ to prepare nearly hexagonal packings in two-dimension (see SI text). We will first show that the granular monolayer effectively postpones the bulk freezing time of the droplet. 

	\section{Freezing delay}
	
	\begin{figure}
		\centering
		\includegraphics[width=8.6cm]{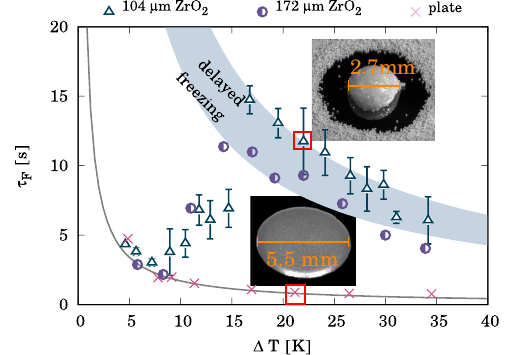}
		\caption{ The freezing time $\tau_F$ of the entire droplet for impacts with the impact velocity $U=1.6\si{m/s}$ on the subfreezing plate ($\times$) and a monolayer of grains ($\triangle$ for $d=104\si{\micro\meter}$ and $\LEFTcircle$ for $d=172\si{\micro\meter}$). For each $\Delta T$, the value of $\tau_F$ is averaged over three measurements. The standard deviation is plotted as error bars for $ d=104\si{\micro\meter} $. The uncertainty for $d=172\si{\micro\meter}$ is comparable, and that for the cold plate is significantly lower (<0.1$\tau_F$). The solid line is a fit of Eq.~\ref{eq.St_freezing}. The two points highlighted by red squares correspond to the solidified morphologies in the insets. The delayed branch remains for higher impact velocities up to \SI{2.75}{m/s}.}
		\label{f.delay}
	\end{figure}

Consider putting a droplet with a volume of $V=D_0^3\pi/6$ into contact with a substrate at the undercooling degree $\Delta T$. The solution to the one-dimensional Stefan problem provides an estimation for the bulk freezing time:
\begin{equation}
	\tau_F=\left(\frac{V}{A}\right)^2\frac{1}{2\alpha_i}\frac{1}{\St}.
	\label{eq.St_freezing}
\end{equation}
In Equation~\ref{eq.St_freezing}, $\alpha_i$ signifies the thermal diffusivity of the solidified phase. The bulk freezing time, $\tau_F$, is inversely related to contact area $A$ and the Stefan number $\St=c_p\Delta T/L$, with $L$ representing the latent heat of the phase transition. 

In Fig.~\ref{f.delay}, we plot $\tau_F$ of the droplet impacting on the bare silicon wafer and the grain layers, respectively, for the impact velocity $ U=1.6\si{m/s} $. For impact on the cold silicone wafer, $\tau_F$ follows Eq.~\ref{eq.St_freezing}. For impact on the grain layer, $\tau_F$ resembles the same relation for low $\Delta T$. However, it begins to rise for $\Delta T>10\si{K}$ and transitions to a separate branch with delayed freezing, increased by an order of magnitude. As all quantities except for $A$ are experimental parameters, we use the measured $\tau_F$ and Eq.~\ref{eq.St_freezing} to estimate the contact area for these two branches. Compared to impacts on the silicon wafer, $A$ is cut to roughly a quarter in the delayed branch, consistent with the observed changes in solidified morphology (Fig.~\ref{f.delay} insets). Given the high wettability of the silicone wafer towards hexadecane, the impacting liquid would wet the substrate upon contact. Therefore, the observed dewetting behavior for $ \Delta T>10\si{K} $ can be attributed to the shielding effect of the grain layer, even though it exhibits wettability towards the impacting liquid as well. This shielding effect must occur at early impact and be sustained during droplet spreading-retraction. The overall spreading-retraction process is measured by the capillary time $ \tau_\gamma = \sqrt{\rho_l D_0^3/\sigma} \approx 15 \si{ms}$, where $\rho_l$ and $\sigma$ are the mass density and surface tension coefficient of the liquid. The freezing delay at seconds is thus an outcome of changes in dynamics at milliseconds. 

	\begin{figure*}
	\centering
	\includegraphics[width=\textwidth]{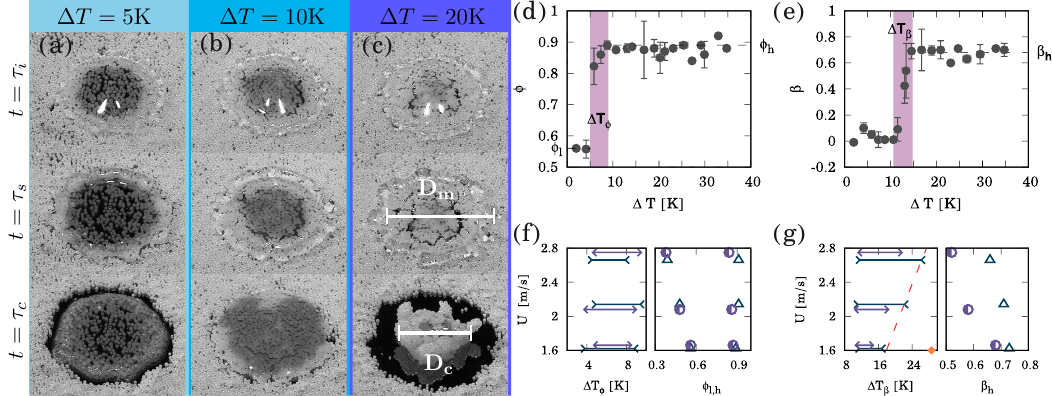}
	\caption{ 
		(a-c): The impact events of $U=1.6\si{m/s}$ on grain layer of $ d=172\si{\micro\meter} $ with various undercooling degrees, $\Delta T$. Snapshots are captured at inertia time $\tau_i$, maximum spreading time $\tau_s$ and retraction time $\tau_c$. In (c), $D_m$ and $D_c$ measure the droplet diameter at the corresponding moments. Panels (d-e) display the particle number ratio $\phi$ and the retraction degree $\beta$ versus $ \Delta T $ for $U=1.6\si{m/s}$ and $d=172\si{\micro\meter}$. Panels (f-g) characterize the variations in $\phi(\Delta T)$ and $\beta(\Delta T)$  with $U$ and $d$, showcasing their lower and upper bounds ($\phi_{l,h}$ and $\beta_h$) and transition regimes ($\Delta T_\phi$ and $\Delta T_\beta$). The same color and symbols as in Fig.~\ref{f.delay} are used for the two grain sizes. In (g), the dashed line represents the theoretical prediction on the upper bound of $\Delta T_\beta$ for the small \ch{ZrO2} grains. The diamond point indicates the critical undercooling temperatures in theory corresponding to Fig.~\ref{f.bottom}d. See text for details. }
	\label{f.impact}
\end{figure*}

\section{Short-time dynamics} 

Figure~\ref{f.impact}a-c show impact events of $U=1.6\si{m/s}$ at three distinct time scales: inertia time $\tau_i$, spreading time $\tau_s$, and retraction time $\tau_c$. The inertia time $\tau_i=D_0/U$ defines the time scale of momentum transfer of the droplet. $\tau_s\approx\tau_\gamma /6$ and $\tau_c\approx 1.5 \tau_\gamma$ respectively marks the endpoint of spreading and retraction (see SI text). Notable differences in dynamics are observable between high and low undercooling degrees. The discrimination occurs earlier as $\Delta T$ increases. Using Fig.~\ref{f.impact}a as a reference, the dark center at $t=\tau_i$ indicates that the liquid permeates through the grain layer and wets the silicon wafer underneath. As time progresses to $t=\tau_s$, the wetted area expands, and the central grains are pushed toward the rim of the spreading liquid film. The droplet then remains in its expanded state with little retraction ($t=\tau_c$) and freezes on the substrate. The bulk freezing time and solidified morphology resemble impacts on a bare silicon wafer substrate [cf Fig.~\ref{f.delay}]. By increasing the undercooling degree to $\Delta T=10\si{K}$, the droplet still wets the central area but without any expansion of the wetting area or movement of the grains within it. For $t > \tau_s$, the liquid-grain mixture exhibits no retraction. However, the liquid in the residual seeps through the inter-grain pores, and the wetting area expands and fully covers the underneath substrate at $t\sim 30\si{ms}\approx 2\tau_\gamma$. For even higher $\Delta T$, the short-time liquid penetration diminishes ($t\leq \tau_i$), and the droplet largely spreads atop the grain layer. Subsequently, pronounced retraction occurs during $\tau_s < t < \tau_c$, and the reduced contact area postpones the bulk freezing [cf. Fig.~\ref{f.delay}].
	
The distinction of impact dynamics depicted in Fig.~\ref{f.impact}a-c is induced by certain freezing processes of the impacting liquid, manifested as the dependence of $\Delta T$. Two dimensionless numbers are introduced to quantify this dependence: $\phi$, the ratio of grain numbers between $t=\tau_s$ and the initial state, and $\beta=(D_{m}-D_c)/(D_m-D_0)$, representing the degree of retraction. In this context, $D_m$ and $ D_c $ denote the droplet diameters at $\tau_s$ and $\tau_c$, respectively (see Fig.~\ref{f.impact}c and SI text for additional details). While calculating $\phi$, grains within the impact center, a square region with a diagonal length of $D_0$, are counted~\footnote{The dimension $D_0$ indicates the span of intense impact stresses that causes most significant particle motion. Other choices, such as the shape of the counting area, does not alter the transition temperature $\Delta T_\phi$}. Figure~\ref{f.impact}d plots $\phi$ for impacts of $U=1.6\si{m/s}$ and $d=172\si{\micro\meter}$. The number ratio transitions from $\phi_l$, at low $\Delta T$, to a plateau, $\phi_h$, at high undercooling degrees. The transition regime is marked as $\Delta T_\phi$. The degree of retraction, $\beta(\Delta T)$, follows a similar pattern (see Fig.~\ref{f.impact}e), undergoing an increase from 0 to its plateau value $\beta_h$ when $\Delta T$ exceeds $\Delta T_\beta$. 

Given the minor contact area between the spherical grains and the silicon wafer and the significant difference in their thermal conductivities, the grains and substrate can be viewed as two separate heat sinks. The freezing processes induced by these two types of heat sinks are referred to as \textit{grain freezing} and \textit{substrate freezing} henceforth. The subsequent analysis will focus on the magnitude and variants of $ \Delta T_{\phi,\,\beta} $, $ \phi_{l,h} $ and $\beta_h$ to clarify the underlying freezing process. 

\subsection{Substrate freezing} 
Grain motion is propelled by the impact stresses of the droplet, measured by the inertia pressure $ \rho_l U^2 $. Consequently, the lower bound $\phi_l$ decreases with increased $U$ as displayed in Fig.~\ref{f.impact}f. In contrast, $\phi_h\approx 0.9$ remains unaffected by $U$, suggesting the formation of a frozen grain patch for $\Delta T>\Delta T_\phi$. The frozen patch needs to form early to counterbalance the impact stresses, preferably within a timeframe comparable to $\tau_i$. As the liquid wets the grain layer and the underneath substrate throughout the transition regime $ \Delta T_\phi $ [cf. Fig.~\ref{f.impact}a-b], both grain freezing and substrate freezing could contribute to the rise of $\phi$. However, their relative significance may yield different outcomes. The former causes the wetted grains to freeze together, while the latter results in the grains freezing onto the substrate (see Fig.~\ref{f.bottom}a and b). In the former case, the grain freezing process covers a dimension proportional to $d$. Therefore, to match the freezing timescale with $\tau_i$, a dependence on $d$ is expected for $\Delta T_\phi$. However, the observed consistency of $\Delta T_\phi$ in Fig.~{\ref{f.impact}f} contradicts this expected relationship, indicating the dominance of substrate freezing for $\phi$ saturation.

To initiate substrate freezing, the droplet must penetrate through the grain layer. Liquid penetration is driven by the inertia pressure of the droplet and is resisted by the viscous drag across a pore dimension $d_p$, $ \sim \mu U/d_p $, with $ \mu $ as the dynamic viscosity of the liquid.  Let $\tau_p$ represent the time required to reach the penetration depth $ d $. It can be shown that $ \tau_p/\tau_i \sim (d/d_p)^2/\Rey$, where $ \Rey=\rho_l U D_0/\mu $ represents the Reynolds number. By measuring the critical impact velocity for $ \tau_p/\tau_i=1 $, we can estimate $ \tau_p $ (see SI text). The duration for the substrate freezing to take effect, $\tau_i-\tau_p$, decreases with $U$ for the parameter range under investigation. Nevertheless, $\Delta T_\phi$ in Fig.~\ref{f.impact}f is insensitive to $U$, which indicates that liquid freezes instantaneously upon contact with the cold substrate. Indeed, the observed upper bound of $\Delta T_\phi$, approximately $9\si{K}$, aligns with the undercooling degree of immediate nucleation for hexadecane reported in Ref.~\citenum{Kant2020}. The area of momentum transfer, equivalent to the area directly beneath the droplet $ \pi D_0^2/4 $ (Fig.~\ref{f.impact}a-b), is the initial penetration site and largely determines the dimension of frozen patch~\footnote{The dimension of the frozen patch is always larger than $D_0$ by approximately 20\% and slightly decreases with $\Delta T$.}.

\subsection{Dewetting initiation} 
Beyond $ \Delta T_\phi $, the droplet body impacts upon the immediately formed frozen layer on the substrate at $ t<\tau_i$. The liquid ejecting from the frozen region primarily possesses horizontal momentum and is isolated from the substrate by the frozen layer, which prompts the dewetting of the spreading lamella and its retraction afterward. As a result, the rise of $\beta$, marked by $\Delta T_\beta$, closely follows the saturation of $\phi$. The lower bound of $\Delta T_\beta$ aligns with the upper bound of $\Delta T_\phi$ within the measurement uncertainty, demonstrating robustness against $U$ and $d$ as well [Fig.~\ref{f.impact}g]. 

Note that the initiation and sustainability of dewetting in this study are largely independent of ambient air~\cite{Grivet2023,Zhao2024}.
Instead, the resistance from grains on the spreading lamella redirects its edge upwards~\cite{Esmaili2021}, similar to droplet impact on a liquid layer~\cite{Peregrine1981,Marcotte2019}. The retraction degree, $\beta$, exhibits significant fluctuations within the transition regime [Fig.~\ref{f.impact}e], primarily due to the lamella recontacting/rewetting the substrate via capillary action. For $\Delta T > \Delta T_\beta$, rewetting diminishes, and the droplet-grain mixture retracts to its smallest feasible size, the frozen patch formed early ($ \sim D_0 $). This defines the saturation value $\beta_h$. The transition regime, $\Delta T_\beta$, aligns with the increase of $\tau_F$ in Fig.~\ref{f.delay}. This confirms the connections between short-time dynamics and the long-term overall freezing.

We assume that a certain separation distance from the substrate, ideally $\sim d/2$, is necessary to ensure the consistent dewetting and the saturation of $\beta$. According to Eq.~\ref{eq.St_freezing}, substrate freezing contributes a frozen dimension of 10\si{\micro\meter} during $\tau_i$ for $U=1.6\si{m/s}$ and $\Delta T=30\si{K}$, approximately one tenth of $d$ or less. Moreover, this frozen dimension remains constant with $d$. Hence, it does not explain the $d$ dependence of $\beta_h$ and the upper bound of $\Delta T_\beta$ in Fig.~\ref{f.impact}g. Therefore, substrate freezing alone cannot saturate $\beta$, necessitating the consideration of grain freezing.	

\subsection{Grain freezing} 

\begin{figure}
	\includegraphics[width=8.6cm]{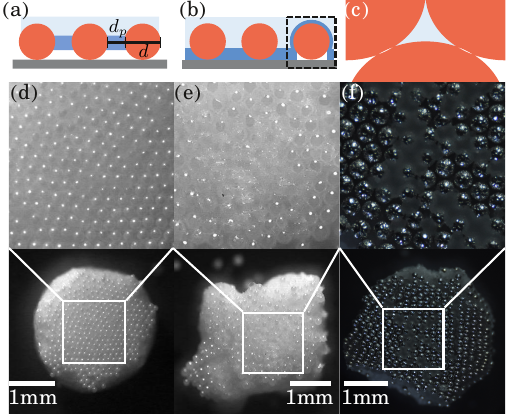}
	\caption{Sketch of grain freezing (a) and substrate freezing (b). The light (dark) blue area indicates the liquid (solidified) phase. The dashed box in (b) illustrates the partial penetration scenario. (c) In a dense monolayer of spheres, the pore (blue area) is a curved triangle. Images of the bottom of the solidified liquid-grain mixture are shown for impacts on small ($d=104\si{\micro\meter}$ in (d)) and large ($d=172\si{\micro\meter}$ in (e)) \ch{ZrO2} layers, and \ch{Sn} balls ($d=155\si{\micro\meter}$) layer in (f). The impact parameters are $U=1.6\si{m/s}$ and $\Delta T= 28\si{K}$ for all cases. The top row gives a zoom-in of the impact center area. Note that the center of the residual is not necessarily the impact center.}
	\label{f.bottom} 
\end{figure}

The process of grain freezing refers to the freezing of liquid that wets the pores in the granular layer. The most pronounced effect of freezing occurs at the smallest cross-section of the pore, as illustrated in Fig.~\ref{f.bottom}c. The grain freezing process is of particular interest beyond $\Delta T_\phi$, where little motion of grains is seen [Fig.~\ref{f.impact}]. A constant pore dimension $d_p = 0.1d$ is thus used in the following analysis~\footnote{This value is obtained by dividing the highlighted area ($(\sqrt{3}-\pi/2)d^2/4$) in Fig.~\ref{f.bottom}c by the arc length of the surrounding grains $d\pi/2$, resulting in a length scale of 0.05$d$. This scale measures the dimension from the grain surface to the pore's center, equivalent to $d_p/2$.}. As freezing occurs from all directions simultaneously, the relevant frozen dimension is $ d_p/2 $. We denote the freezing time over $d_p/2$ as $\tau_{f}$. The classic Stefan problem (Eq.~\ref{eq.St_freezing}) provides $\tau_{f}\approx \tau_i$ at $\Delta T=8\si{K}$ for $U_0=1.6\si{m/s}$ and $d=172\si{\micro\meter}$. This underestimates $\tau_{f}$; otherwise, the significance of grain freezing, such as the variance of $ \Delta T_\phi $ with $ d $, would have been observable. To obtain a more reasonable estimation of $\tau_f$, it is necessary to consider the thermal properties of the heat sink (grains). Two additional dimensionless numbers, $B\equiv \rho_i c_p^i/ \rho_g c_p^g$ and $K\equiv k_i/k_g$, are introduced, representing the ratio of volumetric heat capacity and thermal conductivity between the solidified phase and grains. It is worth noting that the contact area between spherical grains is negligible compared to their surface area. Thus, in the context of grain freezing, the heat sinks are individual grains of a finite dimension $d$. The grain freezing process thus has an inherent length scale and deviates from self-similar solutions like Eq.~\ref{eq.St_freezing}. In Ref.~\citenum{Zhao2024}, we have proposed an approximation solution for $\tau_{f}$ that accounts for the finite size effect to the leading order:
\begin{equation}
	\label{eq.reduced_tauf}
	\tau_{f} = \mathcal{F}(K) \frac{\hat{d}_p^2}{2\St} \left(1+\frac 23\frac B\St \hat{d}_p\right)\frac{d^2}{4\alpha_i},\, \text{with } \hat{d}_p=\frac{d_p}{d}.
\end{equation}
$\mathcal{F}(K) \geq 1$ increases with $K$. We identify $\mathcal{F}(K=0.15)=2$ for \ch{ZrO2} grains (see SI text). The second term in the bracket on the right-hand side of Eq.~\ref{eq.reduced_tauf} is the correction introduced by the finite size effect on the diffusive propagation of the freezing front. This term is about 0.3 at $\Delta T=20\si{K}$ for the material under investigation and decreases for higher $\Delta T$.

Setting $\tau_f=\tau_i$ in Eq.~\ref{eq.reduced_tauf} yields a critical undercooling degree $\Delta T_{i}$. We compute $\Delta T_i$ for the small \ch{ZrO2} grain layer ($d=104\si{\micro\meter}$). The result is approximately inversely proportional to $U$ and aligns with the upper bound of $\Delta T_\beta$ (see the dashed line in Fig.~\ref{f.impact}g). When $\Delta T$ exceeds $\Delta T_i$, $\tau_f$ decreases further, potentially matching the penetration timescale $\tau_p$. The critical undercooling degree that satisfies $\tau_{f}=\tau_p$ is calculated for $U=1.6\si{m/s}$, resulting in $\Delta T_p=28.6\si{K}$. For $\Delta T>\Delta T_p$, the pores are sealed so rapidly by freezing that liquid infiltration halts completely. We show the bottom image of the solidified residual for impacts on the small \ch{ZrO2} layer around $\Delta T_p$. Indeed, grains protrude from the frozen bottom uniformly, with a consistent protrusion height of 11\si{\micro\meter}. Given that $\Delta T_p \sim U^2$, complete suppression of liquid penetration for higher $U$ values was not observed within the studied undercooling range.

The freezing time, $\tau_f$ in Eq.~\ref{eq.reduced_tauf}, is anticipated to exhibit a quadratic increase with the grain size, $d$. Consequently, the larger grains ($d=172\si{\micro\meter}$) requires a significantly greater $\Delta T_{i,p}$ to achieve $\tau_{f}=\tau_{i,p}$, a condition unattainable in our experiments. In other words, grain freezing alone is insufficient to close the pores within $\tau_i$ for the larger grains under the investigated range of $\Delta T$. In this case, consistent dewetting and the saturation of $\beta$ arise from the combined effects of grain freezing and substrate freezing processes, and $\Delta T_\beta$ deviates from the prediction of $\tau_{f}=\tau_i$ by Eq.~\ref{eq.reduced_tauf}. We refer to this scenario as partial penetration. In the partial penetration scenario, the penetrating liquid wets the substrate directly beneath the pore and then spreads. The frozen layer growing from the substrate intersects with that from the grain surface. In consequence, a portion of the grain bottom unwetted [Fig.~\ref{f.bottom}b].
Figure~\ref{f.bottom}e illustrates the bottom of the solidified morphology formed on the large \ch{ZrO2} layer under the same impact and temperature parameters as Fig.~\ref{f.bottom}d, where the impact center is penetrated. The exposure area of grains deviates from the circular shape and increases further away from the impact center. The estimated protrusion height ranges from 0 to 19\si{\micro\meter}, marking the intersection of the frozen layers from the substrate and the grain. Substrate freezing at $\Delta T\sim 30\si{K}$, as previously discussed, fosters the development of a frozen layer approximately 10\si{\micro\meter} thick within $\tau_i$, comparable to the measured protrusion height. The weaker influence of grain freezing for larger grains results in a notable decrease of $\beta_h$ with $U$ (see Fig.~\ref{f.impact}g). The reduced $\beta_h$ indicates a larger frozen area. The freezing delay effect at higher $U$ thus becomes less efficient than that shown in Fig.~\ref{f.delay}.

An increase in thermal conductivity of grains enhances the grain freezing by reducing $ \mathcal{F}(K) $ and thus $\tau_{f}$ in Eq.~\ref{eq.reduced_tauf}. The consequence is demonstrated through an experiment involving a layer of Tin balls with a diameter of $d=155\si{\micro\meter}$, the bottom image of which is presented in Fig.~\ref{f.bottom}f. The high thermal conductivity of Tin ($K\sim 10^{-3}\ll 1$) sets $\mathcal{F}=1$ in Eq.~\ref{eq.reduced_tauf}. From a quantitative standpoint, the square of the diameter ratio between the small \ch{ZrO2} grains and Tin balls, 2.2, matches their difference in $\mathcal{F}$. In other words, the increase in $\tau_{f}$ due to the grain size of Tin balls is offset by their enhanced thermal conductivity. As a result, liquid penetration is significantly inhibited in the Tin ball layer, similar to Fig.~\ref{f.bottom}d. The partial penetration, more visibly apparent here, primarily occurs at the `defects' in the hexagonal structure, i.e., the larger pores.

\section{Conclusion} 

We examine a scenario in this study where the granular layer acts as a shield for the underlying substrate against icing during droplet impact. Compared to superhydrophobic coating, the grain layer configuration investigated here provides an alternative viewpoint. The grain-freezing process could shed the impact stresses (Fig.~\ref{f.impact}d) and the contact of the droplet with the underlying substrate (Fig.~\ref{f.impact}e and \ref{f.bottom}d-f). In consequence, the post-impact residual can be readily removed. For small grains, the Bond number, $\mathrm{Bo} = \rho_g d^2 g/\gamma $, is typically smaller than 1. Therefore, the weight of grains does not affect the droplet retraction. On inclines, weak adhesion that merely balances the gravity can thus be introduced (by static electrical charges, for instance) without altering the droplet retraction and freezing delay effect. After impact, the additional droplet mass overwhelms the weak adhesion, and the residual may be removed under gravity. Lastly, we highlight that the non-wettability of the grain layer could further enhance the shielding effect, as it impedes impact penetration. At the intermediate $\Delta T$ ($\sim \Delta T_\beta$), the non-wettability stabilizes dewetting and retraction. Beyond $\Delta T_p$, however, the dominance of grain freezing reduces the significance of wettability (see SI video).
	
\begin{acknowledgement}
The authors thank Mr. Hao-Jie Zhang for building the Nitrogen chamber and Ms. Yashi Li for her contribution to experimental measurement. This work is supported by NSFC project number 12172277.
\end{acknowledgement} 

\begin{suppinfo}

	The following files are available free of charge.
	\begin{itemize}
		\item SI.pdf: additional details of the experimental protocol, image processing method, the calculation of the penetration time $\tau_p$, and the grain-freezing model. 
		\item SI\_wettability.mp4: illustrating the influence of wettability of grains.
	\end{itemize}
	
\end{suppinfo}

\bibliography{impact_cold_grain}

\providecommand{\latin}[1]{#1}
\makeatletter
\providecommand{\doi}
  {\begingroup\let\do\@makeother\dospecials
  \catcode`\{=1 \catcode`\}=2 \doi@aux}
\providecommand{\doi@aux}[1]{\endgroup\texttt{#1}}
\makeatother
\providecommand*\mcitethebibliography{\thebibliography}
\csname @ifundefined\endcsname{endmcitethebibliography}
  {\let\endmcitethebibliography\endthebibliography}{}
\begin{mcitethebibliography}{19}
\providecommand*\natexlab[1]{#1}
\providecommand*\mciteSetBstSublistMode[1]{}
\providecommand*\mciteSetBstMaxWidthForm[2]{}
\providecommand*\mciteBstWouldAddEndPuncttrue
  {\def\EndOfBibitem{\unskip.}}
\providecommand*\mciteBstWouldAddEndPunctfalse
  {\let\EndOfBibitem\relax}
\providecommand*\mciteSetBstMidEndSepPunct[3]{}
\providecommand*\mciteSetBstSublistLabelBeginEnd[3]{}
\providecommand*\EndOfBibitem{}
\mciteSetBstSublistMode{f}
\mciteSetBstMaxWidthForm{subitem}{(\alph{mcitesubitemcount})}
\mciteSetBstSublistLabelBeginEnd
  {\mcitemaxwidthsubitemform\space}
  {\relax}
  {\relax}

\bibitem[Kreder \latin{et~al.}(2016)Kreder, Alvarenga, Kim, and
  Aizenberg]{Kreder2016}
Kreder,~M.~J.; Alvarenga,~J.; Kim,~P.; Aizenberg,~J. Design of Anti-Icing
  Surfaces: Smooth, Textured or Slippery? \emph{Nature Reviews Materials}
  \textbf{2016}, \emph{1}, 15003\relax
\mciteBstWouldAddEndPuncttrue
\mciteSetBstMidEndSepPunct{\mcitedefaultmidpunct}
{\mcitedefaultendpunct}{\mcitedefaultseppunct}\relax
\EndOfBibitem
\bibitem[Schutzius \latin{et~al.}(2015)Schutzius, Jung, Maitra, Eberle,
  Antonini, Stamatopoulos, and Poulikakos]{Schutzius2015}
Schutzius,~T.~M.; Jung,~S.; Maitra,~T.; Eberle,~P.; Antonini,~C.;
  Stamatopoulos,~C.; Poulikakos,~D. Physics of Icing and Rational Design of
  Surfaces with Extraordinary Icephobicity. \emph{Langmuir : the ACS journal of
  surfaces and colloids} \textbf{2015}, \emph{31}, 4807--4821\relax
\mciteBstWouldAddEndPuncttrue
\mciteSetBstMidEndSepPunct{\mcitedefaultmidpunct}
{\mcitedefaultendpunct}{\mcitedefaultseppunct}\relax
\EndOfBibitem
\bibitem[Jung \latin{et~al.}(2011)Jung, Dorrestijn, Raps, Das, Megaridis, and
  Poulikakos]{Jung2011}
Jung,~S.; Dorrestijn,~M.; Raps,~D.; Das,~A.; Megaridis,~C.~M.; Poulikakos,~D.
  Are {{Superhydrophobic Surfaces Best}} for {{Icephobicity}}? \emph{Langmuir}
  \textbf{2011}, \emph{27}, 3059--3066\relax
\mciteBstWouldAddEndPuncttrue
\mciteSetBstMidEndSepPunct{\mcitedefaultmidpunct}
{\mcitedefaultendpunct}{\mcitedefaultseppunct}\relax
\EndOfBibitem
\bibitem[Guo \latin{et~al.}(2012)Guo, Zheng, Wen, Song, Lin, and
  Jiang]{Guo2012}
Guo,~P.; Zheng,~Y.; Wen,~M.; Song,~C.; Lin,~Y.; Jiang,~L.
  Icephobic/{{Anti-Icing Properties}} of {{Micro}}/{{Nanostructured Surfaces}}.
  \emph{Advanced Materials} \textbf{2012}, \emph{24}, 2642--2648\relax
\mciteBstWouldAddEndPuncttrue
\mciteSetBstMidEndSepPunct{\mcitedefaultmidpunct}
{\mcitedefaultendpunct}{\mcitedefaultseppunct}\relax
\EndOfBibitem
\bibitem[Alizadeh \latin{et~al.}(2012)Alizadeh, Yamada, Li, Shang, Otta, Zhong,
  Ge, Dhinojwala, Conway, Bahadur, Vinciquerra, Stephens, and
  Blohm]{Alizadeh2012}
Alizadeh,~A.; Yamada,~M.; Li,~R.; Shang,~W.; Otta,~S.; Zhong,~S.; Ge,~L.;
  Dhinojwala,~A.; Conway,~K.~R.; Bahadur,~V.; Vinciquerra,~A.~J.; Stephens,~B.;
  Blohm,~M.~L. Dynamics of Ice Nucleation on Water Repellent Surfaces.
  \emph{Langmuir : the ACS journal of surfaces and colloids} \textbf{2012},
  \emph{28}, 3180--3186\relax
\mciteBstWouldAddEndPuncttrue
\mciteSetBstMidEndSepPunct{\mcitedefaultmidpunct}
{\mcitedefaultendpunct}{\mcitedefaultseppunct}\relax
\EndOfBibitem
\bibitem[Maitra \latin{et~al.}(2014)Maitra, Antonini, Tiwari, Mularczyk, Imeri,
  Schoch, and Poulikakos]{Maitra2014}
Maitra,~T.; Antonini,~C.; Tiwari,~M.~K.; Mularczyk,~A.; Imeri,~Z.; Schoch,~P.;
  Poulikakos,~D. Supercooled Water Drops Impacting Superhydrophobic Textures.
  \emph{Langmuir : the ACS journal of surfaces and colloids} \textbf{2014},
  \emph{30}, 10855--10861\relax
\mciteBstWouldAddEndPuncttrue
\mciteSetBstMidEndSepPunct{\mcitedefaultmidpunct}
{\mcitedefaultendpunct}{\mcitedefaultseppunct}\relax
\EndOfBibitem
\bibitem[Varanasi \latin{et~al.}(2010)Varanasi, Deng, Smith, Hsu, and
  Bhate]{Varanasi2010}
Varanasi,~K.~K.; Deng,~T.; Smith,~J.~D.; Hsu,~M.; Bhate,~N. Frost Formation and
  Ice Adhesion on Superhydrophobic Surfaces. \emph{Applied Physics Letters}
  \textbf{2010}, \emph{97}, 234102\relax
\mciteBstWouldAddEndPuncttrue
\mciteSetBstMidEndSepPunct{\mcitedefaultmidpunct}
{\mcitedefaultendpunct}{\mcitedefaultseppunct}\relax
\EndOfBibitem
\bibitem[Kulinich \latin{et~al.}(2011)Kulinich, Farhadi, Nose, and
  Du]{Kulinich2011}
Kulinich,~S.~A.; Farhadi,~S.; Nose,~K.; Du,~X.~W. Superhydrophobic
  {{Surfaces}}: {{Are They Really Ice-Repellent}}? \emph{Langmuir}
  \textbf{2011}, \emph{27}, 25--29\relax
\mciteBstWouldAddEndPuncttrue
\mciteSetBstMidEndSepPunct{\mcitedefaultmidpunct}
{\mcitedefaultendpunct}{\mcitedefaultseppunct}\relax
\EndOfBibitem
\bibitem[Chen \latin{et~al.}(2012)Chen, Liu, He, Li, Cui, Zhang, Zeng, Zhang,
  Wang, and Song]{Chen2012}
Chen,~J.; Liu,~J.; He,~M.; Li,~K.; Cui,~D.; Zhang,~Q.; Zeng,~X.; Zhang,~Y.;
  Wang,~J.; Song,~Y. Superhydrophobic Surfaces Cannot Reduce Ice Adhesion.
  \emph{Applied Physics Letters} \textbf{2012}, \emph{101}, 111603\relax
\mciteBstWouldAddEndPuncttrue
\mciteSetBstMidEndSepPunct{\mcitedefaultmidpunct}
{\mcitedefaultendpunct}{\mcitedefaultseppunct}\relax
\EndOfBibitem
\bibitem[Wang \latin{et~al.}(2019)Wang, Kong, Wang, and Liu]{Wang2019}
Wang,~L.; Kong,~W.; Wang,~F.; Liu,~H. Effect of Nucleation Time on Freezing
  Morphology and Type of a Water Droplet Impacting onto Cold Substrate.
  \emph{International Journal of Heat and Mass Transfer} \textbf{2019},
  \emph{130}, 831--842\relax
\mciteBstWouldAddEndPuncttrue
\mciteSetBstMidEndSepPunct{\mcitedefaultmidpunct}
{\mcitedefaultendpunct}{\mcitedefaultseppunct}\relax
\EndOfBibitem
\bibitem[Schremb \latin{et~al.}(2017)Schremb, Roisman, and Tropea]{Schremb2017}
Schremb,~M.; Roisman,~I.~V.; Tropea,~C. Transient Effects in Ice Nucleation of
  a Water Drop Impacting onto a Cold Substrate. \emph{Physical Review E}
  \textbf{2017}, \emph{95}, 022805\relax
\mciteBstWouldAddEndPuncttrue
\mciteSetBstMidEndSepPunct{\mcitedefaultmidpunct}
{\mcitedefaultendpunct}{\mcitedefaultseppunct}\relax
\EndOfBibitem
\bibitem[Zhang \latin{et~al.}(2020)Zhang, Liu, Wu, and Min]{Zhang2020}
Zhang,~X.; Liu,~X.; Wu,~X.; Min,~J. Impacting-Freezing Dynamics of a
  Supercooled Water Droplet on a Cold Surface: {{Rebound}} and Adhesion.
  \emph{International Journal of Heat and Mass Transfer} \textbf{2020},
  \emph{158}, 119997\relax
\mciteBstWouldAddEndPuncttrue
\mciteSetBstMidEndSepPunct{\mcitedefaultmidpunct}
{\mcitedefaultendpunct}{\mcitedefaultseppunct}\relax
\EndOfBibitem
\bibitem[Zhao \latin{et~al.}(2024)Zhao, Zhang, and Li]{Zhao2024}
Zhao,~S.-C.; Zhang,~H.-J.; Li,~Y. Cold Granular Targets Slow the Bulk Freezing
  of an Impacting Droplet. \emph{Proceedings of the National Academy of
  Sciences} \textbf{2024}, \emph{121}, e2311930121\relax
\mciteBstWouldAddEndPuncttrue
\mciteSetBstMidEndSepPunct{\mcitedefaultmidpunct}
{\mcitedefaultendpunct}{\mcitedefaultseppunct}\relax
\EndOfBibitem
\bibitem[Kant \latin{et~al.}(2020)Kant, Koldeweij, Harth, {van Limbeek}, and
  Lohse]{Kant2020}
Kant,~P.; Koldeweij,~R. B.~J.; Harth,~K.; {van Limbeek},~M. A.~J.; Lohse,~D.
  Fast-Freezing Kinetics inside a Droplet Impacting on a Cold Surface.
  \emph{Proceedings of the National Academy of Sciences} \textbf{2020},
  \emph{117}, 2788--2794\relax
\mciteBstWouldAddEndPuncttrue
\mciteSetBstMidEndSepPunct{\mcitedefaultmidpunct}
{\mcitedefaultendpunct}{\mcitedefaultseppunct}\relax
\EndOfBibitem
\bibitem[Grivet \latin{et~al.}(2023)Grivet, Huerre, S{\'e}on, and
  Josserand]{Grivet2023}
Grivet,~R.; Huerre,~A.; S{\'e}on,~T.; Josserand,~C. Making Superhydrophobic
  Splashes by Surface Cooling. \emph{Physical Review Fluids} \textbf{2023},
  \emph{8}, 063603\relax
\mciteBstWouldAddEndPuncttrue
\mciteSetBstMidEndSepPunct{\mcitedefaultmidpunct}
{\mcitedefaultendpunct}{\mcitedefaultseppunct}\relax
\EndOfBibitem
\bibitem[Esmaili \latin{et~al.}(2021)Esmaili, Chen, Pandey, Kim, Lee, and
  Jung]{Esmaili2021}
Esmaili,~E.; Chen,~Z.-Y.; Pandey,~A.; Kim,~S.; Lee,~S.; Jung,~S. Corona
  Splashing Triggered by a Loose Monolayer of Particles. \emph{Applied Physics
  Letters} \textbf{2021}, \emph{119}, 174103\relax
\mciteBstWouldAddEndPuncttrue
\mciteSetBstMidEndSepPunct{\mcitedefaultmidpunct}
{\mcitedefaultendpunct}{\mcitedefaultseppunct}\relax
\EndOfBibitem
\bibitem[Peregrine(1981)]{Peregrine1981}
Peregrine,~D.~H. The Fascination of Fluid Mechanics. \emph{Journal of Fluid
  Mechanics} \textbf{1981}, \emph{106}, 59--80\relax
\mciteBstWouldAddEndPuncttrue
\mciteSetBstMidEndSepPunct{\mcitedefaultmidpunct}
{\mcitedefaultendpunct}{\mcitedefaultseppunct}\relax
\EndOfBibitem
\bibitem[Marcotte \latin{et~al.}(2019)Marcotte, Michon, S{\'e}on, and
  Josserand]{Marcotte2019}
Marcotte,~F.; Michon,~G.-J.; S{\'e}on,~T.; Josserand,~C. Ejecta, {{Corolla}},
  and {{Splashes}} from {{Drop Impacts}} on {{Viscous Fluids}}. \emph{Physical
  Review Letters} \textbf{2019}, \emph{122}, 014501\relax
\mciteBstWouldAddEndPuncttrue
\mciteSetBstMidEndSepPunct{\mcitedefaultmidpunct}
{\mcitedefaultendpunct}{\mcitedefaultseppunct}\relax
\EndOfBibitem
\end{mcitethebibliography}
\end{document}